\documentclass[aps,prl,floatfix,twocolumn,superscriptaddress,showpacs,epsf]{revtex4-1}

\usepackage{amsmath,amssymb,amsfonts}
\usepackage{epstopdf}
\usepackage{epsfig} 
\usepackage{graphicx}
\usepackage{subfigure}
\usepackage{import}
\usepackage{color}
\usepackage{url}

\allowdisplaybreaks

\usepackage[colorlinks=true,linkcolor=blue,citecolor=red]{hyperref}

\makeatletter
\renewcommand\tableofcontents{\@starttoc{toc}}
\makeatother
\def\bcen{\begin{center}}
\def\ecen{\end{center}}

\def\e{\varepsilon}          
              \def\n{\nu}

\def\aa{{\V \a}}

\def\=={\equiv}

\def\qed{\raise1pt\hbox{\vrule height5pt width5pt depth0pt}}

\def\cG0{{\cal G}_0} 
\def\cG{{\cal G}}

\def\bk{{\bf k}}
\def\bq{{\bf q}}

 \def\=={\equiv}

 \def\ep0{\epsilon_{p}} \def\ed0{\epsilon_{f}}

\def\be{\begin{equation}}
\def\ee{\end{equation}}

\def\cc{c^{\dagger}}
\def\ca{c^{\phantom{\dagger}}}

\def\ac{a^{\dagger}}
\def\aa{a^{\phantom{\dagger}}}

\newcommand{\ket}[1]{|{#1}\rangle}

\newcommand{\braket}[3]{\langle{#1}| {#2} |{#3} \rangle}
\newcommand{\quave}[1]{\langle {#1} \rangle}

\newcommand{\bd}[1]{\mathbf{#1}}

\usepackage{fancyhdr}
\begin{document}
\author{Giacomo Mazza}
\email{giacomo.mazza@polytechnique.edu}
\affiliation{CPHT, Ecole Polytechnique, CNRS, Universit\'e Paris-Saclay, 91128 Palaiseau, France}
\affiliation{Coll\`ege de France, 11 place Marcelin Berthelot, 75005 Paris, France}
\author{Antoine Georges}
\affiliation{Coll\`ege de France, 11 place Marcelin Berthelot, 75005 Paris, France}
\affiliation{Center for Computational Quantum Physics, Flatiron Institute,  
162 Fifth avenue, New York, NY 10010, USA} 
\affiliation{CPHT, Ecole Polytechnique, CNRS, Universit\'e Paris-Saclay, 91128 Palaiseau, France}
\affiliation{DQMP, Universit\'e de Gen\`eve, 24 quai Ernest Ansermet, CH-1211 Gen\`eve, Suisse}

\title{Superradiant Quantum Materials}

\begin{abstract}
  There is currently great interest in the strong coupling between the 
  quantized photon field of a cavity and electronic or other degrees of freedom in materials. 
  A major goal is the creation of novel collective states entangling photons with those degrees of freedom. 
  Here we show that the cooperative effect between strong electron correlations in quantum materials and 
  the long-range interactions induced by the photon field leads to the stabilization 
  of coherent phases of light and matter.  
  By studying a two-band model of interacting electrons coupled to a cavity field, we show that 
  a phase characterized by the simultaneous condensation of excitons and photon superradiance 
  can be realized, hence stabilizing and intertwining two collective phenomena which are rather 
  elusive in the absence of this cooperative effect. 
\end{abstract}

\maketitle

\paragraph{Introduction.}
Collective phenomena due to interactions between light and matter have become in 
recent years a major focus of interest spanning different fields of research. 
By allowing to create and control entangled quantum states of light and matter, 
cavity quantum electrodynamics (QED) offers a fascinating platform in this context. 
This has led to several highly successful research directions, in fields as diverse as 
atomic physics~\cite{Haroche_book,haroche_RMP,Haroche_quantum_field,wineland_RMP},
quantum information~\cite{wineland_logic_gate,mabuchi_quantum_network,kimble_quantum_computation,hartmann_cavity_array} 
and quantum fluids of polaritons~\cite{amo_polariton_superfluids,amo_solitons_polariton_superfluid,ciuti_carusotto_RMP}. 
Advances in controlling and probing light-matter interactions have allowed for the investigation of 
collective effects in solid-state systems such as atomically thin or layered materials~\cite{basov_polaritons_vdw,dufferwiel_vdw, 
mak_qed_2dtmd,liu_MoS2_nat_phot,xia_2d_nanophotonics,liu_prl_WS2,splendiani_MoS2,mak_MoS2,xia_2d_nanophotonics,mak_qed_2dtmd}.
Recently, pioneering work has also explored strong light-matter coupling in
molecules and molecular solids~\cite{ebbesen_perspective,hutchison_merocyanine,schwartz_merocyanine}.

One of the earliest and most important examples of collective phenomena in 
coupled light-matter systems is \emph{superradiance}.
Originally introduced by Dicke \cite{dicke_original} in the description of the 
collective enhancement of spontaneous emission, superradiance 
signals a coherence in a ensemble of dipoles 
collectively interacting with the same radiation field.
At equilibrium superradiance appears as phase 
transition~\cite{hepp_superradiance,wang_superradiance}
characterized by the macroscopic population
of photons in the ground state and the collective 
ordering of dipoles induced by a photon-mediated effective 
dipole-dipole interaction.
Analogous transitions have been studied in non-equilibrium 
conditions~\cite{baumann_dicke_bec,dalla_torre_keldysh_dicke} 
and in the context of circuit QED~\cite{ciuti_nogo,Viehmann_prl_SR,ciuti_comment,bamba_sr_circuitQED,jaako_pra_2016}.


In condensed matter physics, a wealth of emergent collective phases has been found in 
`quantum materials'~\cite{tokura_quantum_materials}, which result from strong
interactions between electrons as well as other degrees of freedom.
In this context, the engineering of new forms of effective interactions 
by means of collective light-matter coupling raises the fascinating possibility 
of exploring novel emergent collective phenomena 
involving entangled states of light and matter~\cite{laussy_sc_cavity,imagoglu_sc_cavity,smolka_many_body_2d,sentef_arxiv_cavity,schlawin_arxiv_cavity}. 
\begin{figure}[b]
  \includegraphics[width=0.925\linewidth]{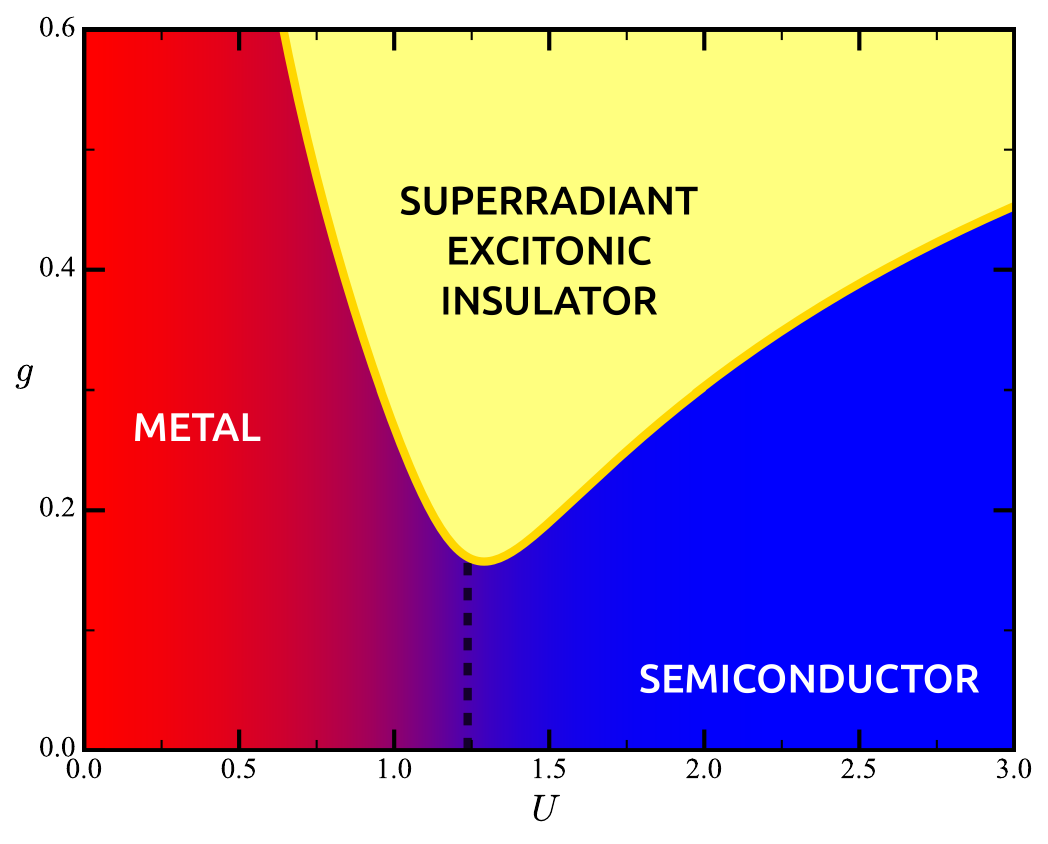}
  \caption{
    Light-matter phases as a function of the strength of electronic
    interactions $U$ and the light-matter coupling $g$ at a temperature $T=0.15$ 
    The red/blue intensity map reflects the evolution of the band populations. 
    In the metal region (red) both orbitals are occupied, whereas in the semiconductor
    region (blue) the valence band is  completely filled.
  }
  \label{fig1}
\end{figure}

Littlewood and Zhu proposed early on~\cite{littlewood_ei} that superradiance 
may occur when electron-hole transitions in semiconducting quantum wells are
coupled to a cavity photon. 
In their work, electronic interactions were not included and, most importantly, 
the photon diamagnetic term, which has been shown to 
impede superradiance~\cite{nogo_PRL_1975,ciuti_nogo,Viehmann_prl_SR}, was neglected. 

In this Letter, we demonstrate theoretically in a simple model of interacting electrons
coupled to a cavity field how a cooperative effect between light-matter coupling and intrinsic
electronic interactions can lead to the stabilization of a coherent light-matter phase, the `superradiant excitonic
insulator' (SXI), characterized by the simultaneous appearance 
of equilibrium superradiance in the photon
field and the condensation of excitons in the electronic system. 
As summarized in Fig.~\ref{fig1}, superradiance cannot be reached in the absence of
electronic interactions. At the same time coupling to the cavity promotes excitonic
condensation in regimes in which it cannot be stabilized by electronic
interactions only.
Hence, SXI is a phase that intertwines superradiance and excitonic condensation
in conditions where the two phases cannot be individually stabilized.

\paragraph{Light-matter Hamiltonian.}
We consider spinless electrons moving in valence ($\nu=1$) and
conduction ($\nu=2$)  bands with hopping parameters $t_2=-t_1=t_{hop}$, 
originating from localized atomic orbitals separated by an energy gap  $\omega_{12}$.
The electrons interact via a local repulsive interaction $U$ acting when two electrons sit on the
same site. The electronic Hamiltonian reads: 
\begin{equation}
  H_{el} = \sum_{\bk \nu} \e_{\nu}(\bk) \cc_{\bk \nu} \ca_{\bk \nu} +
  U  \sum_{i} n_{i 1} n_{i 2} - \mu \sum_{i \nu} n_{i \nu}
  \label{eq:H_excitonic_ins}
\end{equation}
where $\cc_{\bk \nu}/\ca_{\bk \nu}$ is the creation/annihilation operator
for  an electron in the Bloch state $\ket{\bk \nu}$ with
quasi-momentum $\bk$ in the band $\nu$ and $n_{i \nu} = \cc_{i \nu}
\ca_{i \nu}$ is the electron number operator on lattice site $i$.
While the results do not depend qualitatively on this choice,  
we consider for simplicity a one-dimensional lattice. 
The  dispersion relations read
$\e_{2}(\bk) = -\e_{1}(\bk) = \frac{\omega_{12}}{2} - 2 t_{hop} \cos{k}$ and
we choose $\mu=\frac{U}{2}$ 
to fix the density to one electron per site $\quave{n_{i1}+n_{i2}} = 1$

The coupling of the electronic system (\ref{eq:H_excitonic_ins}) 
to an optical cavity is described
by the light-matter Hamiltonian:
\begin{eqnarray} \nonumber
H&=&H_{el} + \omega_0 \ac\aa +\Delta \hat{\rho} \left( a+ \ac \right)^2 + \\
&+&\left(a+\ac\right) \frac{1}{\sqrt{N}}\sum_{\bk} \sum_{\nu \nu'} 
g_{\nu \nu'}(\bk)\cc_{\bk \nu} \ca_{\bk \nu'}
\label{eq:Hk_light_matter}
\end{eqnarray}
corresponding to a single-mode of frequency $\omega_0$ with vector potential
$\bd{A}(\bd{r}) \approx \frac{\bd{A}_0}{\sqrt{N}} \left( a+ a^{\dagger} \right)$
(dipole approximation) polarized along the spatial dimension of the electronic system.
The third term in (\ref{eq:Hk_light_matter})
is the diamagnetic contribution with $\Delta = \frac{e^2}{2m} \bd{A}_0^2$ and $\bd{A}_0 =
\sqrt{\frac{\rho}{2\omega_0 \epsilon_0}} \bd{u} $, 
while the last term is the $\bk-$dependent dipolar coupling between $\ket{\bk \nu}$
Bloch states, $g_{\nu \nu'}(\bk) = \frac{e}{m} \braket{\bk \nu}{ \bd{p}}{\bk \nu'} \cdot \bd{A}_0$. 
$\rho=N/V$ is the electronic density with $V$ the cavity volume.
$\epsilon_0$ is the permittivity of the cavity and $\bd{u}$ the polarization vector.

The couplings of the diamagnetic and dipolar terms are not independent:
they are related by the sum rules resulting from the canonical commutation
relations $i=\left[ \bd{r},\bd{p} \right]$~\cite{ciuti_nogo,Viehmann_prl_SR}.
In the Bloch basis we find that for inter-band transitions
the following relation holds for each $\bk$
\begin{equation}
  \Delta =\sum_{\n \ne \n'} \frac{|g_{\n \n'}(\bk)|^2}{\varepsilon_{
      \n'} (\bk) - \varepsilon_{\n} (\bk)}.
  \label{eq:gk_sum}
\end{equation}
We assume to assign all the oscillators strength to the transition between the
two low-energy bands $\nu=1,2$, obtaining
\begin{equation}
  g_{12}(\bk) = g f_{\bk} 
  \label{eq:lm_coupling}
\end{equation}
where  $g \equiv \sqrt{\Delta~\omega_{12}}$ is the light-matter coupling
with dimension of an energy and $f^2_{\bk} \equiv | \e_1(\bk) - \e_2(\bk) |  / \omega_{12}$
is a dimensionless factor  characterizing the momentum dispersion.
We neglect the intra-band couplings as they do not play any role
for the phase transition discussed here.
The nature and validity of the approximations made 
are summarised in details in the Supplemental
Material~\cite{supplementary}.

In the limit 
$U = t_{hop} = 0$ 
the Hamiltonian (\ref{eq:Hk_light_matter}) reduces to the well known Dicke-Hopfield
model of localized dipoles~\cite{dicke_original,hopfield_model}.
Dipole-dipole interactions between localized dipoles have been considered in~\cite{keeling_coulomb_dicke,vukics_Asquare,bamba_polarizable}.
For finite hopping, but $U=0$,  the model has been recently employed 
to describe charge transport in cavity-coupled 
semiconductors~\cite{orgiu_organic_semiconductor,hagenmuller_cavity_transport}.
Energies are measured with respect to $\omega_{12} = 1$ and we fix $\omega_{12}=\omega_0$. 

The Hamiltonian (\ref{eq:Hk_light_matter}) has a global continuous
symmetry 
$\ca_{i \nu} \to e^{i \varphi_{\nu}} \ca_{i \nu}$ and 
$\aa \to e^{i \lambda} \aa$ with $\varphi_1-\varphi_2 = \lambda=\pm \pi$.
We introduce two order parameters:  
the macroscopic expectation value of the photon field $\alpha\equiv\langle a \rangle/\sqrt{N}$
signaling superradiance~\cite{wang_superradiance,hepp_superradiance}, 
and $\Phi \equiv \langle \cc_{i1} \ca_{i2} \rangle$ associated with the condensation of particle-hole pairs (excitons). 
The latter opens an hybridization gap,  inducing an insulating phase known as an `excitonic insulator'
(EI)~\cite{kozlov_divalent_crystal,kohn_excitonic_ins,keldysh_collective_properties,kunes_review}.   
A non-zero value of either $\alpha$ or $\Phi$ breaks the above symmetry.

\begin{figure*}
  \includegraphics[width=0.925\linewidth]{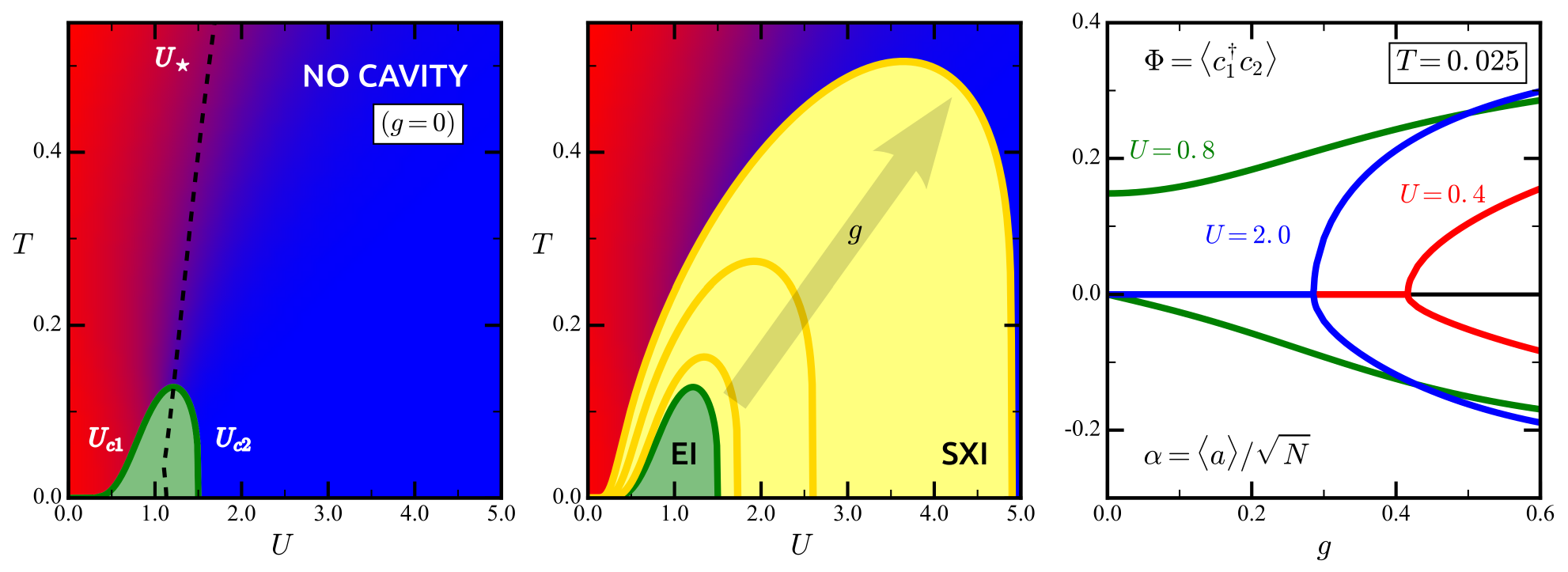}
  \caption{
    (a) Phase diagram in the absence of light-matter coupling: metal(red), semiconductor (blue)
    and excitonic insulator (green).
    $U_{c1}$ and $U_{c2}$: critical interactions for the EI.
    The red/blue intensity map: orbital polarization  as in Fig.~\ref{fig1}.
    Dashed line: critical interaction for gap opening in the normal phase.
    (b) Phase diagram for increasing  $g$ (arrow): $g=0$ (green line) 
    $g=0.2,~0.4~\text{and}~0.6$ (yellow lines). 
    At finite $g$ the EI (green) transforms into a SXI (yellow). 
    (c) Excitonic and superradiant order parameters as a function of $g$ at $T=0.025$ and 
    three values of $U$.
  }
  \label{fig2}
\end{figure*}

\begin{figure}[b]
  \includegraphics[width=0.925\linewidth]{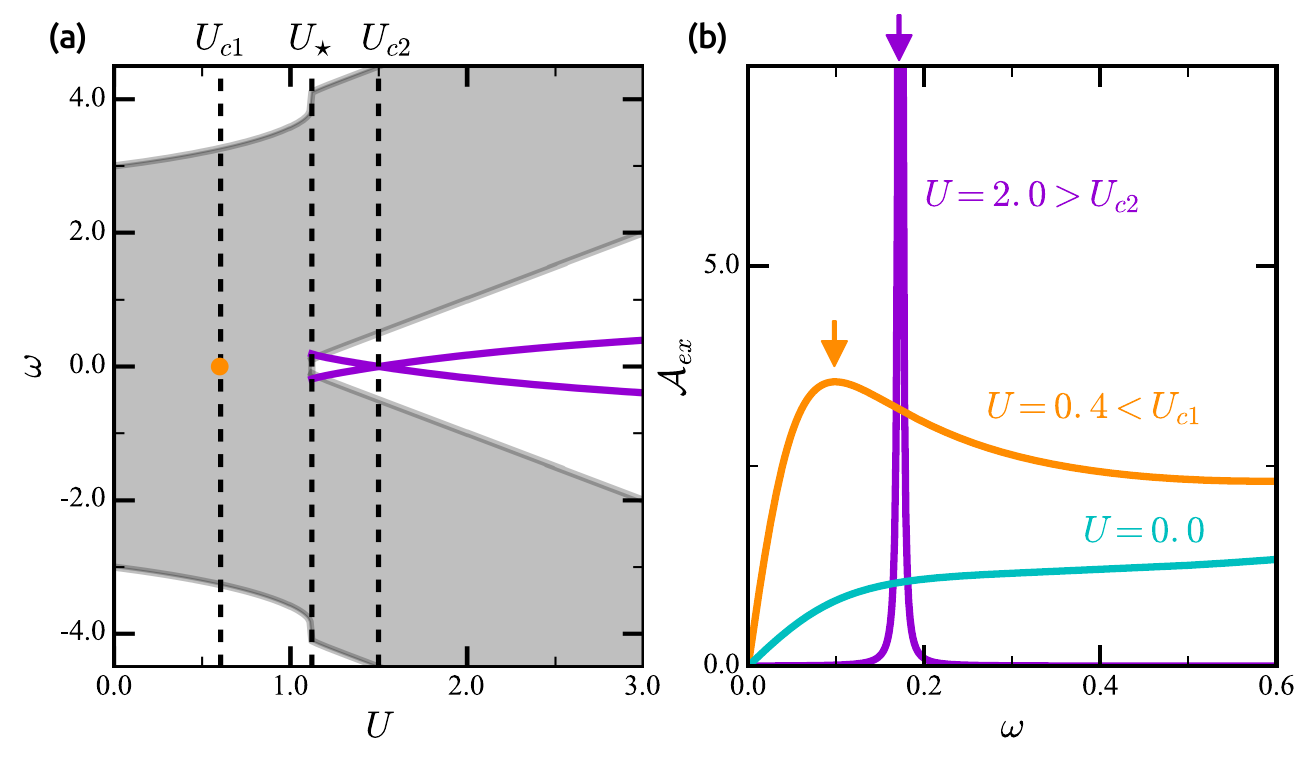}
  \caption{
    (a) Spectrum of  particle-holes excitations as a function of $U$ at $T=0.025$ at $g=0$.
    The shaded area represents the particle-hole continuum. Dashed lines: $U_{c1}$ and $U_{c2}$
    and $U_\star$ as defined in Fig.~\ref{fig2}a.
    The dot at $U=U_{c1}$ indicates a $\omega=0$ pole.
    Full lines: branches of poles crossing zero at $U=U_{c2}$.
    (b) Density of particle-hole excitations for three values of $U$.
    Arrows: excitonic resonances.
  }
  \label{fig3}
\end{figure}

In the thermodynamic limit for the electronic system, 
the light matter interaction can be treated exactly as
the saddle point condition 
of an effective extensive action for the photon field
\begin{equation}
 \Omega_0\, \alpha + 
\frac{1}{N} 
\sum_{\bk} \Gamma(\bk) \left[  \quave{\cc_{\bk
    1} \ca_{\bk 2}}_{\alpha}  + \quave{\cc_{\bk  2} \ca_{\bk 1}}_{\alpha} \right] = 0
\label{eq:saddle_point}
\end{equation}
where $\Omega_0 \equiv \sqrt{\omega_0^2 + 4 \Delta\omega_0}$ and
$\Gamma(\bk) \equiv  g_{12}(\bk) \sqrt{\frac{\omega_0}{\Omega_0}}$.
The electronic averages $\quave{\cdots}_{\alpha}$  have to be
computed with an effective interacting Hamiltonian
$ H_{el}^{\mathrm{eff}} = H_{el} + 2 \alpha \sum_{\bk} (\Gamma(\bk) \cc_{\bk 1} \ca_{\bk 2} + h.c.) $
which is treated by introducing a Hartree-Fock (HF) decoupling of the interaction term
$n_{i1} n_{i 2} \to -m\,(n_{i1}-n_{i2}) - \Phi\,\cc_{i 2} \ca_{i 1}- \Phi^* \cc_{i 1} \ca_{i 2} +\mathrm{const}$,
where $m \equiv \quave{n_{i1}} - \quave{n_{i2}} $ is the electronic orbital polarization.

\paragraph{Absence of superradiance.}
The critical light-matter coupling $g_c$ for superradiance is found
by solving (\ref{eq:saddle_point}) for $g$ in the limit $\alpha \to 0^+$. 
In the case of non-interacting electrons $U=0$ at zero temperature this corresponds to the solution of 
\begin{equation}
  \Omega_0 - \frac{4}{N}\sum_{\bk} \frac{\Gamma(\bk)^2}{|\varepsilon_{1}(\bk) - \varepsilon_{2}(\bk)|} = 0.
  \label{eq:saddle_point_V0}
\end{equation}
Using the definition of $\Omega_0$ and $\Gamma(\bk)$
it is easy to see that this condition can never be satisfied for any value of $g$,
electronic dispersions $\varepsilon_{\nu} (\bk)$ or $\omega_0 > 0$.
This shows that no superradiant transition is possible: it is prevented  
by the diamagnetic coupling $\Delta$ which grows as $g$ is increased, Eq.~(\ref{eq:lm_coupling}).
This result extends to the case of itinerant electrons the so called `no-go theorem'
for the Dicke transition~\cite{nogo_PRL_1975,ciuti_nogo,Viehmann_prl_SR}.
The same result is obtained in the opposite atomic limit $t_{hop} \to 0$ 
independently of  
electronic interactions~\cite{birula_nogo_1979,gawedzki_no_go_1981}.

\begin{figure*}
  \includegraphics[width=0.925\linewidth]{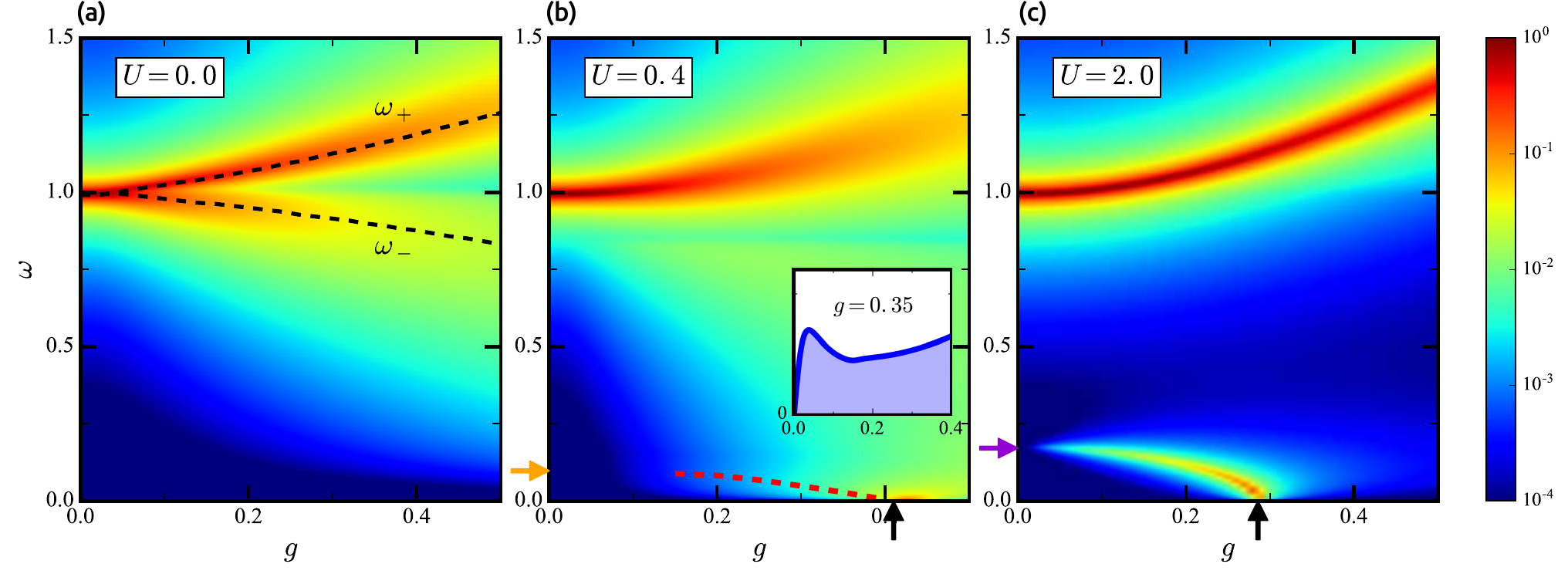}
  \caption{
    Photon spectral functions ${\cal A}_{ph}$ as a function of the light-matter
    coupling $g$ for $U=0$ (a), $U=0.4$ (b) and $U=2.0$ (c) at $T=0.025$.
    Dashed lines in panel (a): the upper and lower polaritons.
    Dashed line in panel (b): dispersion of the low-frequency resonance in the
    spectral function (see inset).
    Vertical arrows in panels (b) and (c) indicate the critical
    couplings for the SXI. Horizontal arrows: excitonic resonances at $g=0$, as defined in
    Fig.~\ref{fig3}b.
  }
  \label{fig4}
\end{figure*}

\paragraph{Superradiant excitonic insulator.}
The above picture dramatically changes once both
electronic interactions $U$ and electronic delocalization $t_{hop}$
are taken into account. From now on we fix $t_{hop} = 0.5$.
For $g=0$, Fig.~\ref{fig2}(a),
the electronic system is unstable towards the formation of an EI below a dome-shaped critical temperature
$T_c(U)$~\cite{kozlov_divalent_crystal,kohn_excitonic_ins,keldysh_collective_properties,kunes_review}, 
between two critical interactions $U_{c1}(T)<U<U_{c2}(T)$ that merge
at the maximum $T_{c,0}^{max}$.
At high-temperature the system evolves from a metal ($0<m<1$) for $U<U_\star$
to a direct gap semiconductor ($m \simeq 1$) for $U>U_\star$
with a gap opening up between the two effective HF bands at $U=U_\star$.

At finite light-matter coupling, as expected by the above symmetry considerations, 
the EI phase transforms into a superradiant excitonic insulator (SXI), characterized by
non-zero superradiant and excitonic order parameters (panel (c)).
This is easily understood as the EI ground state is characterized by a 
macroscopic electronic dipole moment~\cite{theory_electronic_ferrolectricity,supplementary}.
Remarkably, when the system is not in the EI phase at $g=0$ 
there is (for $U\neq 0$) a critical value of the  light-matter coupling $g_c$
beyond which the SXI appears. 
As a result the phase space (temperature and
interactions) for which the SXI is realized is significantly enhanced
as the strength of the light-matter coupling is increased (Fig.~\ref{fig2}(b)).

The occurrence of this intertwined light-matter collective phase is 
shown on Fig.~\ref{fig1}, displaying the phase diagram as a function $U$ and $g$ 
for a temperature at which no coherent phase can be stabilized in the absence of
light-matter coupling.
The SXI phase  intrudes between the metallic and semiconducting phases. 
At weak interaction $U<U_\star$,  $g_c$ increases upon decreasing $U$ 
and diverges as $U \to 0$, as expected from the no-go theorem discussed above. 
On the contrary, in the strong interaction regime, $g_c$ is an increasing 
function of $U$ and approaches the finite value
$g_c^{\infty} = \frac{1}{2} \sqrt{\frac{\omega_0 \omega_{12}}{ f_{loc}^2 - 1}}$
with $f_{loc} = \frac{1}{N} \sum_{\bk} f_{\bk}$ as $U\to \infty$.
At intermediate values of the interaction, corresponding to $U \sim U_\star$,
the critical coupling has a minimum.
By decreasing temperature such a dip in the phase boundary moves towards
the $g=0$ axis until it becomes zero for $T=T_{c,0}^{max}$ and splits into  
two points at $U=U_{c1}$ and $U=U_{c2}$ respectively for $T<T_{c,0}^{max}$.

\paragraph{Exciton-polariton softening.}
The entangled nature of the SXI phase is investigated by 
looking at polariton modes resulting from the dressing of the 
cavity photon with the electronic transitions.
This is characterized by the excitonic susceptibility
$\chi(\omega) = \sum_{\bk \bk'} \chi^{12}_{\bk \bk'} (\omega) + \chi^{21}_{\bk \bk'} (\omega)$.
Here,  $\chi^{\nu \nu'}_{\bk \bk'}(\tau-\tau') = -\quave{{\cal T}_{\tau}
  \cc_{\bk \nu}(\tau)\ca_{\bk \nu'}(\tau) \cc_{\bk' \nu'}(\tau')\ca_{\bk' \nu}(\tau')}$ 
  are the two-particles Green's functions computed in the Random Phase
Approximation (RPA) and in the absence of light-matter coupling:
\begin{equation}
  \chi_{\bk \bk'}(\omega) = \delta_{\bk \bk'} \chi^{0}_{\bk}(\omega) 
  - \frac{U}{N}\chi^{0}_{\bk}(\omega) \sum_{\bq} \chi_{\bq \bk'}(\omega)
  \label{eq:RPA_chi}
\end{equation}
and $\chi^{0 }_{\bk} (\omega)$ the bare ($U=0$) susceptibility.

The spectrum of particle-hole excitations as a function
of the interaction $U$ is displayed on Fig.~\ref{fig3}.
At $U=0$ the spectrum is characterized by a featureless
particle-hole continuum. At finite $U$ we observe the formation of
well defined excitonic modes.
At weak interaction $U < U_{\star}$ this mode corresponds to a resonance embedded
in the particle-hole continuum, while for $U >U_{\star}$, an
exciton gets split-off from the continuum and becomes a sharp pole inside the
semiconducting gap~\cite{zenker_bcsbec}.

At finite light-matter coupling, the particle-hole excitations
hybridize with the photon through the polarization
$\Pi(\omega) = \frac{1}{N} \sum_{\bk \bk'} \Gamma(\bk) \Gamma(\bk') \left[\chi^{12}_{\bk \bk'} (\omega) + \chi^{21}_{\bk \bk'} (\omega)\right]$.
This gives rise to the dressed photon spectra
${\cal A}_{ph}(\omega) $ displayed in Fig.~\ref{fig4} as a function of $g$
for several values of $U$.
In the non-interacting case $U=0$ (panel (a)), the dressing produces
two (lower and upper) polariton branches $\omega_{\pm}$ originating from the bare photon resonance.
At finite interaction $U$, the lower branch shifts to low frequency,
becoming a finite-width resonance for $U<U_\star$ (panel (b)) 
and a sharp mode for $U>U_\star$  (panel (c)). Both modes become
soft at the critical coupling $g_c$ for the SXI transition.

At $U=0$ the polariton modes result from the dressing by
the continuum of bare particle-hole excitations. The latter are constrained
by the sum-rules which prevent the SXI transition and, in turn, the photon softening.
On the contrary, at finite $U$, the lower polariton modes originate from the correlated
excitonic modes discussed above (horizontal arrows in panels (b) and (c)).
Therefore, in the interacting case, the photon couples to an excitation of the
many-body system and, as a result, the sum-rule for the bare particle-hole excitations
no longer prevents a superradiant state  which can indeed  be reached at a finite critical
value of the light-matter coupling.

The cooperation between the electronic interactions and
light-matter coupling in the formation of the SXI state
can be rationalized on the basis of a Landau expansion of the free-energy in terms of the two linearly
coupled ordered parameters $\alpha$ and $\Phi$, which reads:
$F[\alpha,\Phi]=a_\alpha \alpha^2 + a_\Phi \Phi^2 + 2k\, \alpha \Phi + b\Phi^4 + \cdots$
with $a_{\alpha} > 0$ and $a_\Phi = c (T-T_c^0)$, with $T_c^0$ the critical temperature associated
with the uncoupled EI phase. At finite $k$ an instability related to
a linear combination of the excitonic and superradiant eigenmodes occurs for $a_\alpha a_\Phi < k^2$,
so that the SXI phase can be stabilized for $T>T_c^0$ for $k^2>k_c^2=c(T-T_c^0)a_\alpha$
yielding an enhanced critical temperature $ T_c=T_c^0 +k^2/ca_\alpha$.

The SXI transition does not rely on the pumping of photons into the cavity~\cite{keeling_polariton_condensation_short,keeling_polariton_condensation_long}.
The coupling with an external environment, neglected here, is expected 
to bring only quantitative changes to
the above physics such as shifting the critical couplings or 
temperatures~\cite{dalla_torre_keldysh_dicke,scarlatella_dissipative_SR}.
However, this could play a crucial role in the observation of the SXI, in particular for 
the detection of the superradiant ground state~\cite{ciuti_carusotto_input_output}
or the observation of the soft polariton modes~\cite{keller_critical_softening}.

In conclusion, we have investigated the cooperation between
collective light-matter coupling and intrinsic electronic correlation. This
leads to the SXI phase, entangling superradiance and excitonic condensation.
Our results draw attention to quantum materials with strong electronic
correlations as ideal test beds for the observation of entangled quantum states
of light and matter.
Specifically, collective light-matter coupling could be used as a probe 
for  excitonic condensation in systems with a potential excitonic 
instability~\cite{exp_ei_TiSe2,kogar_TiSe2,seki_TNS}. 
In this respect, we emphasize that recent experimental investigations 
of Ti$_2$NiSe$_5$ have suggested exciton condensation~\cite{seki_TNS,disalvo_Ta2NiSe5,exp_ei_Ta2NiSe5,TNS_ultrathin_film,TNS_fano_resonance} 
in this material, and that optical excitations have revealed the interaction 
of the excitonic condensate with light~\cite{mor_ufast_ei,murakami_photoi_ei,kaiser_TNS}.
Other possible candidates for exciton condensation providing a promising 
testing grounds for the SXI transition are electron-hole bi-layers~\cite{review_bilayers} such as bi-layer 
graphene~\cite{mac_donald_nat_phys,mac_donald_exciton_bilayer_graphene,zhang_exciton_bilayer_graphene,cao_TBG_MI,cao_TBG_SC}. 
%

\begin{acknowledgments}
  We acknowledge useful discussions with  T.~Ebbesen, G.~Pupillo, A.~Rubio, 
  C.~Ciuti and F.~Schlawin. 
  This work has been supported by the European Research Council (ERC-319286-`QMAC'). 
  The Flatiron Institute is supported by the Simons Foundation (AG).
\end{acknowledgments}

\bibliography{biblio_cavities}

\end{document}